\documentclass[lettersize,journal]{IEEEtran}
\usepackage{amsmath,amsfonts}
\usepackage{algorithmic}
\usepackage{algorithm}
\usepackage{array}
\usepackage[caption=false,font=normalsize,labelfont=sf,textfont=sf]{subfig}
\usepackage{textcomp}
\usepackage{stfloats}
\usepackage{url}
\usepackage{verbatim}
\usepackage{graphicx}
\usepackage{cite}

\DeclareRobustCommand{\IEEEauthorrefmark}[1]{\smash{\textsuperscript{\footnotesize #1}}}

\begin{document}

\title{Development of a microwave SQUID multiplexer for magnetic microbolometers}

\author{\IEEEauthorblockN{\textbf{
		N. Müller\IEEEauthorrefmark{1,2,3,4},
		J. Bonilla-Neira\IEEEauthorrefmark{1,2,4,6},
		J. Geria\IEEEauthorrefmark{1,2,4,6},
		M. García Redondo\IEEEauthorrefmark{1,3,4,5},
		L. Ferreyro\IEEEauthorrefmark{1,3,4,6},
        M. Hampel\IEEEauthorrefmark{1,3},
        M. Wegner\IEEEauthorrefmark{5,2},
		A. Almela\IEEEauthorrefmark{1,3} and
		S. Kempf\IEEEauthorrefmark{2,5}}}

\IEEEauthorrefmark{1}Instituto de Tecnologías en Detección y Astropartículas (ITeDA), Argentina\\
\IEEEauthorrefmark{2}Institute of Micro- and Nanoelectronic Systems (IMS), Karlsruhe Institute of Technology (KIT), Germany\\
\IEEEauthorrefmark{3}Comisión Nacional de Energía Atómica (CNEA), Argentina\\
\IEEEauthorrefmark{4}Universidad Nacional de San Martín (UNSAM), Argentina\\
\IEEEauthorrefmark{5}Institute for Data Processing and Electronics (IPE), Karlsruhe Institute of Technology (KIT), Germany\\
\IEEEauthorrefmark{6}Consejo Nacional de Investigaciones Científicas y Técnicas (CONICET), Argentina\\
Email: nahuel.muller@iteda.gob.ar}

\maketitle

\begin{abstract}
The search for primordial B-modes in the cosmic microwave background (CMB) requires highly sensitive and scalable detector systems. The magnetic microbolometer (MMB) is an emerging detector concept that exploits the magnetic properties of paramagnetic materials at sub-kelvin temperatures, offering bolometers with a high dynamic range and low intrinsic noise. In recent years, the microwave SQUID multiplexer ($\mu$MUX) has become a key technology to efficiently read large low temperature detector arrays, enabling the readout of hundreds to thousands of detectors over a single transmission line with low noise and minimal power dissipation while reducing the cryogenic setup complexity. In this work, we report the design, fabrication and characterization of a $\mu$MUX optimized for MMB detectors and share our latest experimental results from a bolometer prototype. These findings provide valuable insight of the $\mu$MUX in advancing next-generation CMB instrumentation and also demonstrate its suitability for novel detector technologies such as the MMB.
\end{abstract}


\section{Introduction}
The discovery of B-modes in the polarization patterns of the cosmic microwave background would provide direct evidence for an inflationary epoch in the early Universe \cite{Spergel1997}. Achieving this goal requires detector technologies capable of resolving extremely faint signals, while remaining scalable to the large arrays demanded by next-generation experiments. Within this context, our group is developing the magnetic microbolometer (MMB) \cite{Geria2023}, a recent novel cryogenic bolometer type that is based on the well-established magnetic microcalorimeter (MMC) technology.

\begin{figure}[!h]
\centering
\includegraphics[width=\columnwidth]{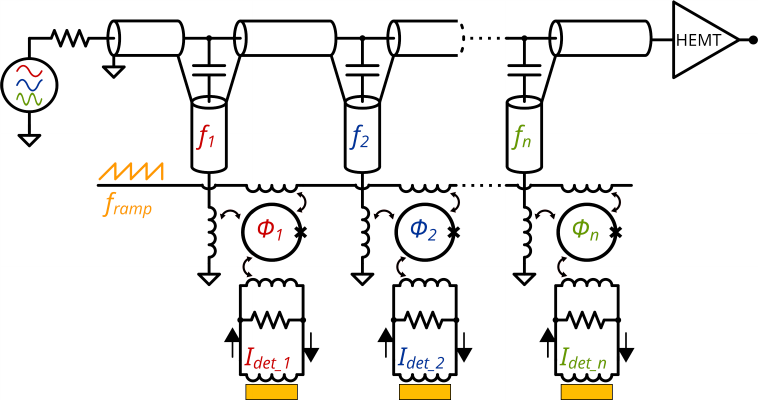}
\caption{Schematic of a microwave SQUID multiplexer with $N$ channels. Each quarter-wave microwave resonator ($f_\mathrm{1}, f_\mathrm{2}, \dots, f_\mathrm{n}$) is coupled to a common feedline and magnetically linked to an rf-SQUID, which shifts its resonance frequency in response to the input flux ($\Phi_\mathrm{1}, \Phi_\mathrm{2}, \dots, \Phi_\mathrm{n}$). The curved arrows indicate the inductive couplings to the rf-SQUID.}
\label{nam_squid_umux}
\end{figure}

To read out these bolometers, we propose the use of the Microwave Superconducting Quantum Interference Device Multiplexer ($\mu$MUX) \cite{Irwin2004}. This device is widely used with MMCs and transition edge sensors (TES), enabling multiplexing factors in the order of thousands of detectors \cite{Kempf2017,Wegner2018,Dober2021,Hirayama2013,Mates2017}. The $\mu$MUX is a GHz-range frequency-division multiplexing (FDM) device consisting of a single transmission feedline capacitively coupled to $N$ microwave resonators, each with a well-defined and unique characteristic frequency (Fig.~\ref{nam_squid_umux}). Each resonator is inductively coupled to a non-hysteretic rf-SQUID, which transduces a flux change in its input coil into a resonator frequency shift. In addition, all rf-SQUIDs within the device are coupled to a common modulation line, that is used to linearize their response through flux-ramp modulation \cite{Mates2012,Salum2023}.

\section{Device fabrication}
The multiplexer devices were fabricated at the Institute of Micro- and Nanoelectronic Systems (IMS) at Karlsruhe Institute of Technology (KIT). The process begins with a Nb/Al-AlOx/Nb trilayer ($100\,\mathrm{nm}/7\,\mathrm{nm}/100\,\mathrm{nm}$) that serves as the foundation for the Josephson tunnel junctions (JJs). The coplanar waveguide (CPW) feedline and resonators, along with the rf-SQUID loops and modulation line, were patterned in the bottom Nb layer using inductively coupled plasma reactive ion etching (ICP-RIE). Silicon dioxide ($\mathrm{SiO_2}$) isolation layers were then deposited in two consecutive steps using RF magnetron sputtering to minimize pinhole formation. A resistive AuPd layer was then sputtered and patterned to implement the RL low-pass filters in the detector input circuit. Finally, the JJs counter-electrode and the remaining wiring were defined by lift-off on a sputtered $450\,\mathrm{nm}$ Nb layer.

\begin{figure}[!t]
\centering
\includegraphics[width=\columnwidth]{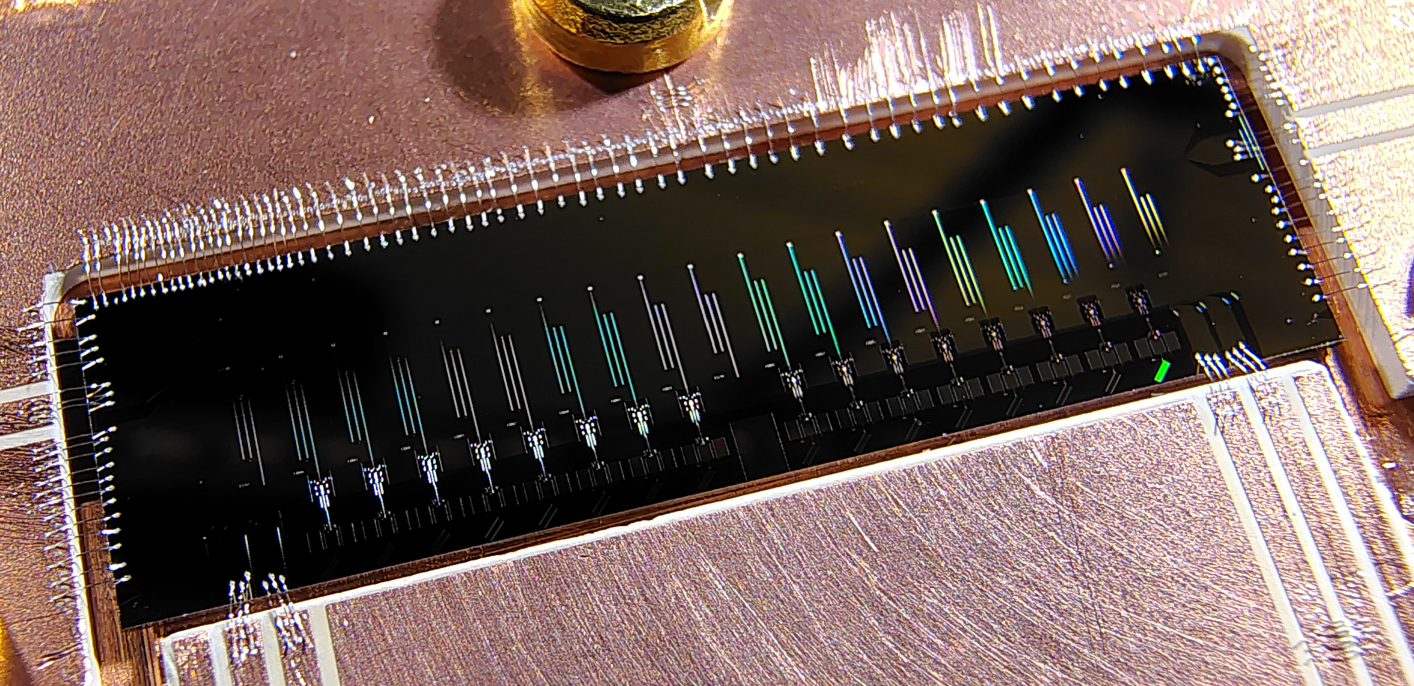}
\vskip 0.25em
\includegraphics[width=0.65\columnwidth]{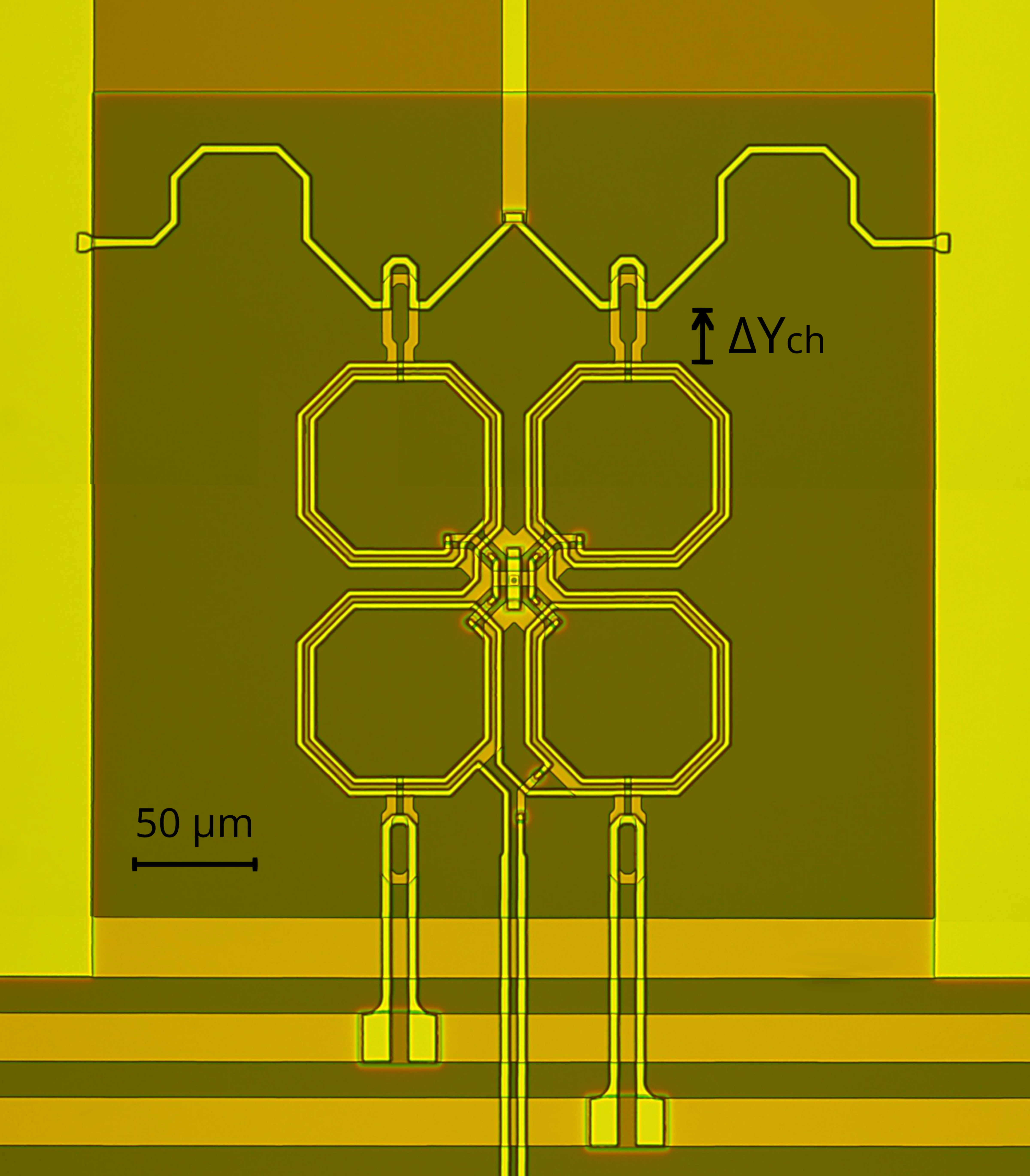}
\caption{(Top) Sixteen-channel SQUID multiplexer with a $20 \times 5\,\mathrm{mm}^2$ die size, mounted on a copper housing for characterization. Three additional bare CPW resonators were included to verify intra-chip fabrication parameters. (Bottom) Microscope image of a four-lobe gradiometric rf-SQUID. The JJ lies in the middle of the design, with the modulation line and resonator couplings at the bottom and top, respectively. The latter is tuned for each channel by adjusting the distance $\Delta Y_\mathrm{ch}$ between the resonator and the rf-SQUID.}
\label{NM_TMUX_03_bonded_NM_QUBIC_03_zoom}
\end{figure}

We fabricated devices with 8, 16, and 32 channels, covering the frequency range of our room-temperature electronics ($4 - 8\,\mathrm{GHz}$) \cite{Gartmann2024}. In this work, we focus on results obtained with the 16-channel device (Fig.~\ref{NM_TMUX_03_bonded_NM_QUBIC_03_zoom}). The characteristic channel frequencies $f_0$ were defined by the length of the quarter-wave resonators and were arranged to maximize the separation between neighboring frequencies. The corresponding design parameters are summarized in Table~\ref{table_parameters}, and their performance was simulated using the $\mu$MUX power-dependence model described in \cite{Wegner2022}. To enable scaling toward 1000 channels within the $4 - 8\,\mathrm{GHz}$ band, the resonator bandwidth was set to $200\,\mathrm{kHz}$, ensuring sufficient channel spacing.

\begin{table}[!h]
\renewcommand{\arraystretch}{1.5}
\caption{$\mu$MUX design parameters}
\centering
\begin{tabular}{l c c}
\hline
Resonator bandwidth & $BW_\mathrm{res}$ & $200\,\mathrm{kHz}$ \\
Channel frequencies & $f_\mathrm{0}$ & $5.744-6.256\,\mathrm{GHz}$ \\
Channel spacing & $\Delta f_\mathrm{0}$ & $32\,\mathrm{MHz}$ \\
Resonance shift & $\Delta f_\mathrm{r}$ & $200\,\mathrm{kHz}$ \\
SQUID-Input coupling & $M_\mathrm{SI}$ & $316\,\mathrm{pH}$ \\
SQUID-Resonator coupling & $M_\mathrm{ST}$ & $2.58-2.34\,\mathrm{pH}$ \\
SQUID-Modulation coupling & $M_\mathrm{SM}$ & $10\,\mathrm{pH}$ \\
\hline
\end{tabular}
\label{table_parameters}
\end{table}

Each rf-SQUID consists of a four-lobe gradiometric loop with a $3.5 \times 3.5\,\mathrm{\mu m^2}$ JJ at its center. The design process resulted in a self-screening parameter $\beta_\mathrm{L} = 2\pi I_\mathrm{C} L_\mathrm{S} / \Phi_\mathrm{0} = 0.5$, selected to prevent hysteretic behaviour. Unlike the TES-SQUID scheme, here the rf-SQUID input inductance $L_\mathrm{IN}$ and the MMB coil $L_\mathrm{DET}$ form a closed superconducting loop (Fig.~1), which partially screens the magnetic coupling $M_\mathrm{TS}$ between the resonator and rf-SQUID \cite{Kempf2017}. This effect was included in the design, leading to a predicted $\sim$12\% deviation between nominal and effective parameters:

$$M_\mathrm{ST_\mathrm{\mathrm{eff}}} = M_\mathrm{ST_\mathrm{\mathrm{design}}} - \frac{M_\mathrm{SI} \, M_\mathrm{TI}}{L_\mathrm{IN} + L_\mathrm{DET}}.$$

The coupling SQUID-Resonator $M_\mathrm{ST}$ was individually adjusted for each channel by shifting the rf-SQUID position relative to the termination inductor $L_\mathrm{T}$. This geometry sets the maximum resonance shift $\eta = \Delta f_\mathrm{r} / BW_\mathrm{\mathrm{res}}$, which was designed aiming to match $\eta = 1$ at an equivalent RF drive power $\Phi_\mathrm{RF} \approx 0.3\,\Phi_0$, consistent with the predictions made in \cite{Schuster2023}.

\begin{figure}[!h]
\centering
\includegraphics[width=0.6\columnwidth]{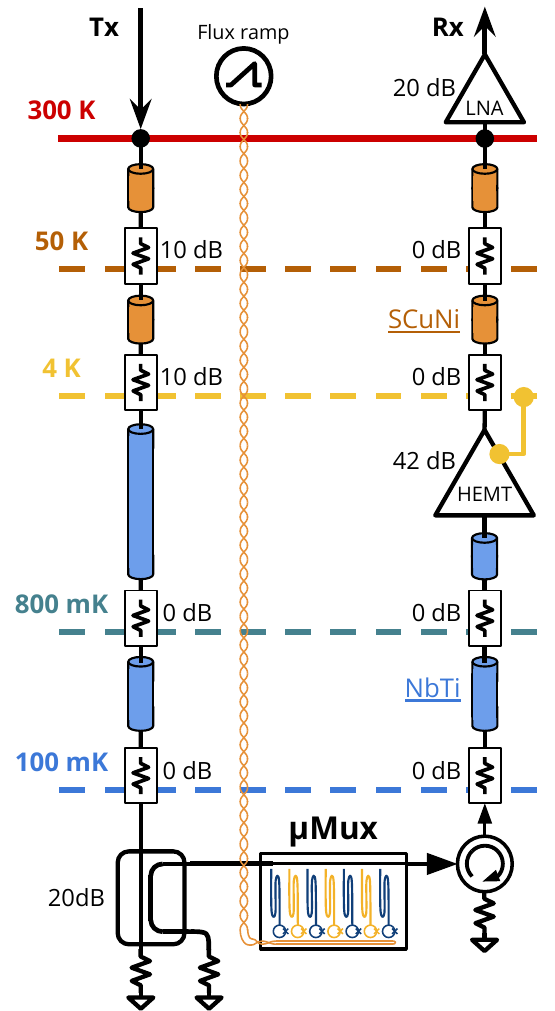}
\caption{Schematic of the experimental setup. The RF attenuations are indicated for each stage of the Bluefors LD250 dilution cryostat and the different coaxial lines materials are specified in blue (Niobium-titanium) and orange (Silver plated copper–nickel). The $\mu$MUX is located in the MXC at a controlled temperature of $100\,\mathrm{mK}$.}
\label{2025.07-uMux_RF_setup}
\end{figure}

\section{Prototype characterization}
The cryogenic measurement setup is shown in Fig.~\ref{2025.07-uMux_RF_setup}. A 16-channel $\mu$MUX device was mounted in a shielded copper sample holder on the mixing chamber (MXC) stage of a Bluefors LD250 dilution refrigerator, stabilized at $T_\mathrm{MXC} = 100\,\mathrm{mK}$. The input transmission line $T_\mathrm{X}$ was attenuated by 40 dB using fixed attenuators distributed across the different temperature stages. On the output ($R_\mathrm{X}$) branch, a total gain of approximately $62\,\mathrm{dB}$ was provided by a Low Noise Factory LNF-LNC4\_8C high-electron-mobility transistor (HEMT) amplifier at $4\,\mathrm{K}$, followed by a Mini-Circuits ZX60-83LN12 low-noise amplifier (LNA) at room temperature. The modulation current was injected via low-ohmic twisted-pair cabling.

\begin{figure}[!h]
\centering
\includegraphics[width=\columnwidth]{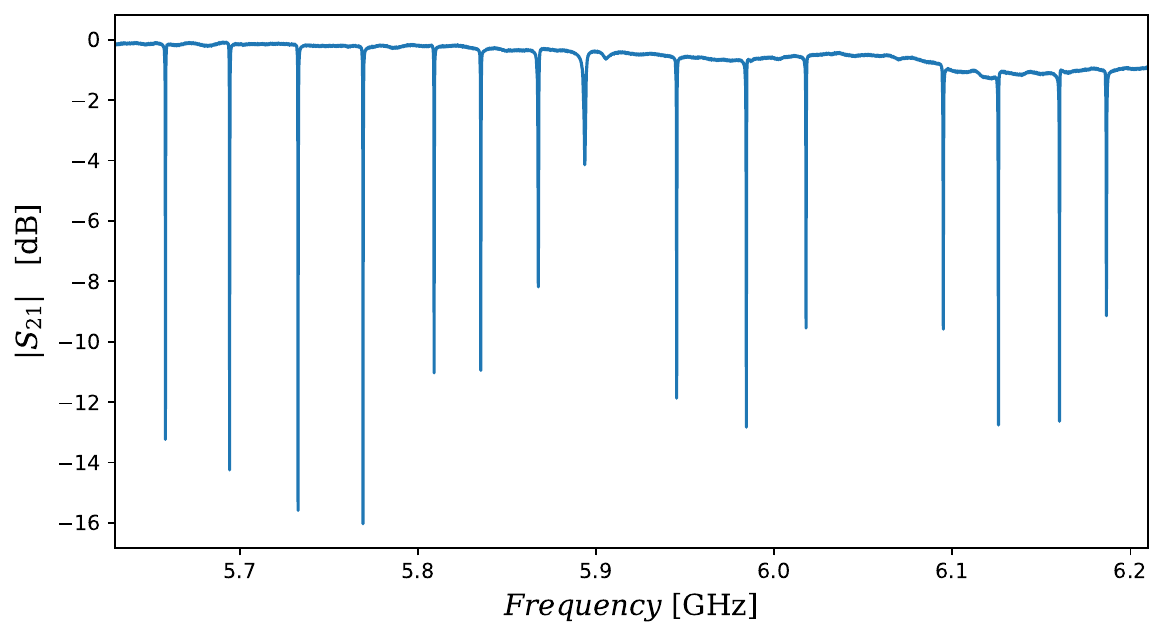}
\caption{$\mu$MUX transmission parameter amplitude $|S_\mathrm{21}|$. This batch included 16 channels separated by $32\,\mathrm{MHz}$ in two $224\,\mathrm{MHz}$ bands of 8 resonators. There is one missing channel due to defects produced during the fabrication process.}
\label{NM_QUBIC_03_S21}
\end{figure}

Readout tone generation, transmission, and demodulation were performed with a custom ZCU216-based software-defined radio (SDR) \cite{Redondo2024,Gartmann2022}. Fig.~\ref{NM_QUBIC_03_S21} shows the transmission parameter amplitude $|S_\mathrm{21}|$ of the 16 channels, with the exception of one channel at a frequency $f \sim 6.05\,\mathrm{GHz}$ due to defects on the microfabrication process. Resonators were grouped in two sets of eight, separated by $64\,\mathrm{MHz}$ to match the SDR local oscillator. The measured characteristic frequencies were shifted by approximately 1.25\% from the design values.

\begin{figure}[!h]
\centering
\includegraphics[width=\columnwidth]{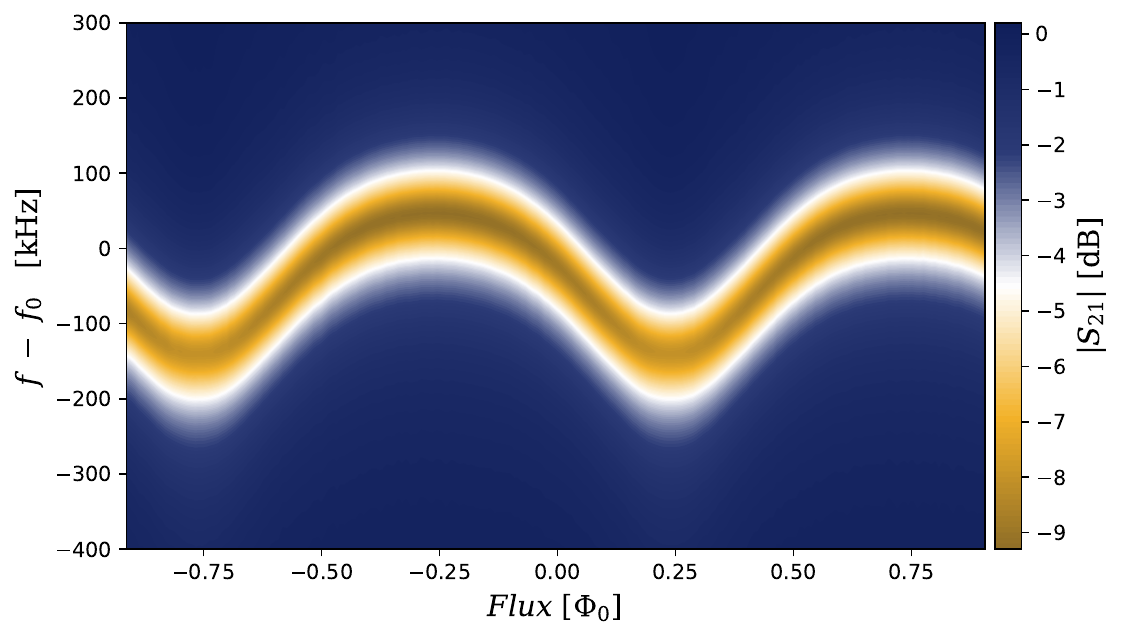}
\caption{Typical $\mu$MUX channel transmission parameter amplitude $|S_\mathrm{21}|$ as a function of input flux and frequency detuning from the resonance $f_\mathrm{0}$. The test flux was applied through the modulation line.}
\label{NM_TMUX_03_S21_fit}
\end{figure}

\begin{figure}[!h]
\centering
\includegraphics[width=\columnwidth]{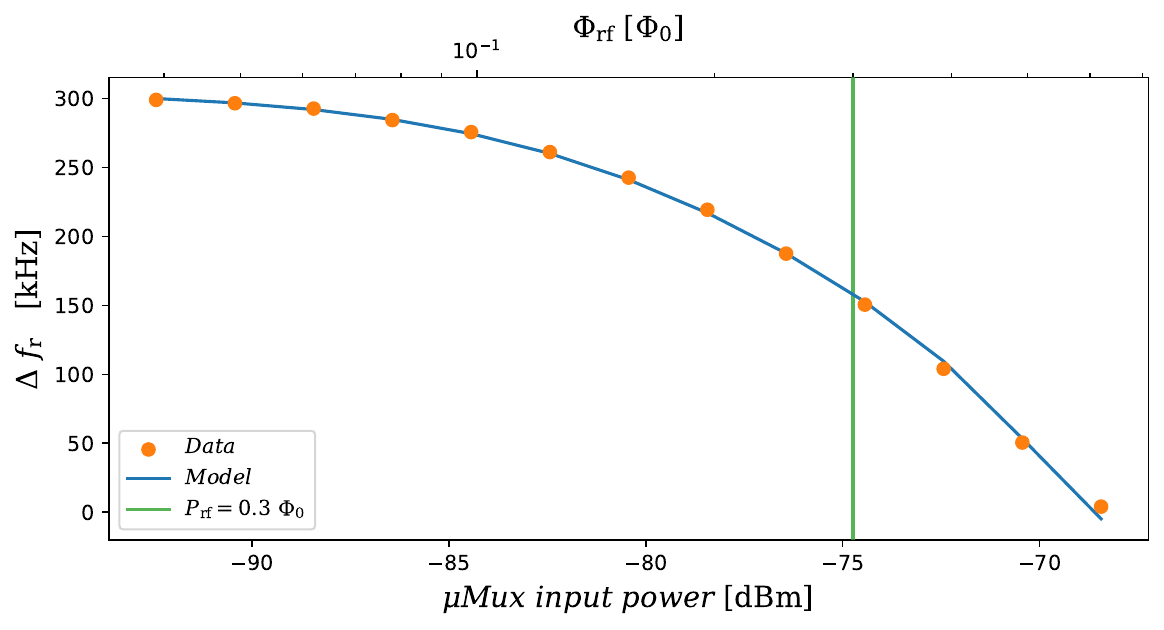}
\caption{Typical $\mu$MUX channel resonance maximum frequency shift as a function of RF readout tone power. The vertical green line indicates the optimal readout power $\Phi_\mathrm{RF} = 0.3\,\Phi_\mathrm{0}$ that was targeted for the shift parameter $\eta = 1$ during the design.}
\label{NM_TMUX_03_S21_power}
\end{figure}

The periodic rf-SQUID response shown in Fig.~\ref{NM_TMUX_03_S21_fit} was obtained by injecting fixed currents through the modulation line. This measurement was repeated for varying RF readout tone powers to characterize the expected power dependence of the resonance-shift parameter (Fig.~\ref{NM_TMUX_03_S21_power}).

\section{MMB prototype: Magnetization measurements}
\begin{figure}[!h]
\centering
\includegraphics[width=0.95\columnwidth]{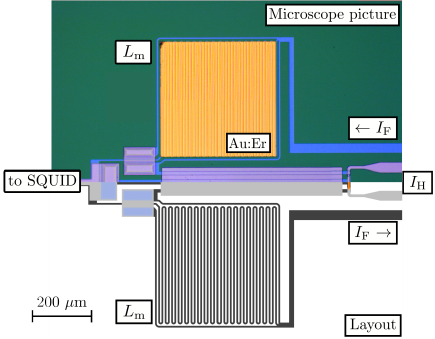}
\caption{MMB prototype microscope picture (top) and corresponding layout (bottom). The symmetric meander-shaped coils consist of a top coil coated with Au:Er sensor material and a bottom coil serving as the load inductor. The heat switch current ($I_\mathrm{H}$) and the persistent bias field current ($I_\mathrm{F}$) injection path are indicated.}
\label{MagTest_MT_ChipDesign}
\end{figure}

To demonstrate the functionality of our $\mu$MUX, we measured flux–temperature $\Phi(T)$ curves of a MMB prototype under development. The sample consisted of two identical gradiometric meander-shaped coils: one used as a load inductor and the other as a pickup coil. The latter is coated with the paramagnetic sensor alloy Au:Er, as shown in Fig.~\ref{MagTest_MT_ChipDesign}.

Two of these devices were connected to different channels of the previously characterized $\mu$MUX. Using the setup described above, a probe RF tone was applied to each channel at a frequency $f_\mathrm{exc}$ above the maximum resonance-shifted value. The corresponding tone amplitudes were recorded while sweeping the cryostat temperature from $500\,\mathrm{mK}$ down to $20\,\mathrm{mK}$. This measurement procedure was repeated for increasing values of the bias field current $I_F$.

To extract $\Phi(T)$ from the data, the raw output signals were first corrected for the temperature-dependent frequency shift of the resonators \cite{Gao2008}, using the zero-bias field data as a reference, and then normalized.

\begin{figure}[!h]
\centering
\includegraphics[width=\columnwidth]{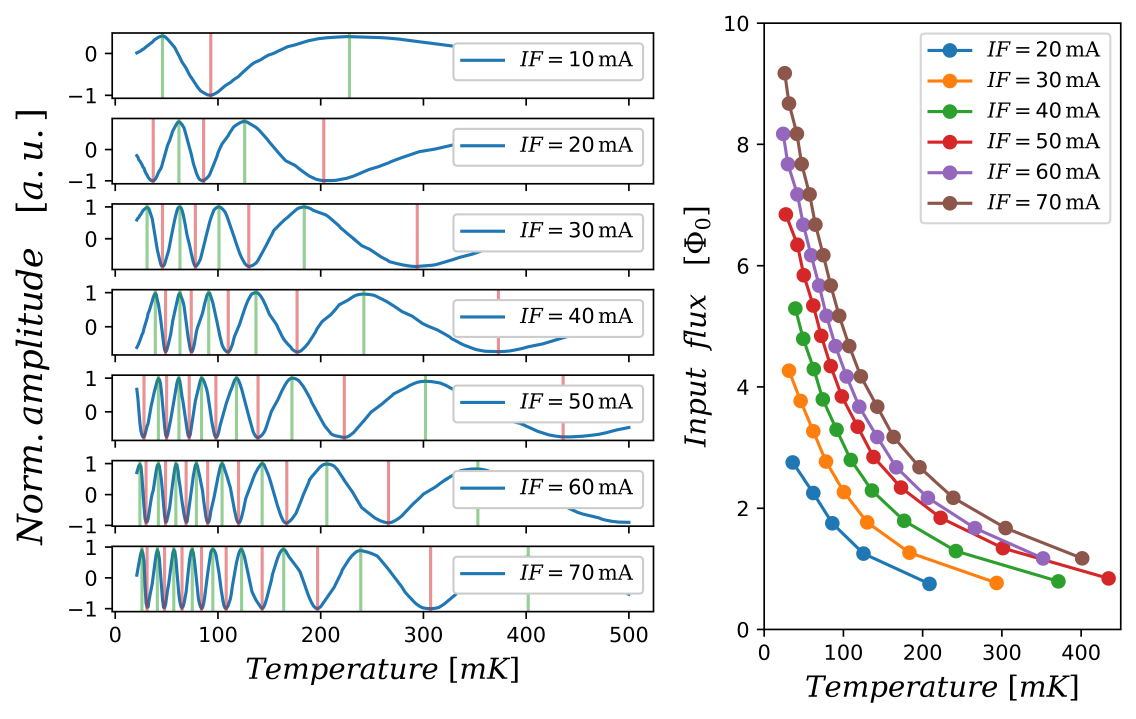}
\caption{(Left) Normalized amplitude of a probe RF tone through a $\mu$MUX channel with a paramagnetic sample in its input as function of temperature for multiple bias field current. The maxima and minima of the rf-SQUID periodic response are indicated with red and green. (Right) Extracted characteristic magnetization curves of the MMB prototype.}
\label{MagTest_merge}
\end{figure}

As shown in left side of Fig.~\ref{MagTest_merge}, the resulting curves exhibit the expected periodic SQUID response, with successive maxima and minima corresponding to changes of $0.5\,\Phi_\mathrm{0}$ in the flux coupled into the input coil by the paramagnetic sensor. The obtained $\Phi(T)$, shown in right side of Fig.~\ref{MagTest_merge}, are consistent with the expected simulated values in \cite{Geria2023}.

\section{Conclusion}
We have reported the design, fabrication, and cryogenic characterization of a microwave SQUID multiplexer optimized for the readout of magnetic microbolometers. A 16-channel $\mu$MUX device was successfully tested at sub-kelvin temperatures, showing stable performance and resonance shifts consistent with the expected RF power dependence. As a proof of principle, prototype MMB detectors were coupled to the multiplexer, and flux–temperature curves were extracted, enabling direct probing of the magnetization of the paramagnetic sensor material using the $\mu$MUX. Future work will focus on detailed noise characterization and the realization of a fully integrated readout chain with functional MMB arrays.

\section*{Acknowledgment}
N. Müller, J. Bonilla-Neira and J. Geria acknowledge support by the Helmholtz International Research School in Astroparticles and Enabling Technologies (HIRSAP) and the Karlsruhe School of Elementary and Astroparticle Physics: Science and Technology (KSETA). This work was partially funded by the Deutsche Forschungsgemeinschaft (DFG, German Research Foundation) – Projektnummer (project number) 467785074.

\bibliographystyle{IEEEtran}
\bibliography{mybib}

\end{document}